\documentstyle[12pt,epsf,rotate]{article}
\begin{document}
\begin{sloppypar}
\begin{center}
{\bf PHASE DIAGRAM FOR CHARGE DENSITY WAVES IN\\
 A  MAGNETIC FIELD}

\vspace{5mm}

D. Zanchi$^1$, A. Bjeli\v{s}$^2$ and G. Montambaux$^1$

\vspace{5mm}

{\em $^1$Laboratoire de Physique des Solides, associ\'{e} au CNRS\\
Universit\'e Paris-Sud, 91405 Orsay, France\\

\vspace{5mm}

$^2$Department of Physics, Faculty of Science,\\
University of Zagreb, POB 162, 41001 Zagreb, Croatia\\}

\vspace{5mm}

ABSTRACT\\

\end{center}
{\small

 The influence of an external magnetic field on a quasi one-dimensional
system    with a charge density wave (CDW) instability is treated within
the random phase approximation (RPA) which includes both CDW and spin
density wave (SDW) correlations. We show that the  CDW
is sensitive to both
 orbital and Pauli effects of the
field. In the case of  perfect nesting, the critical temperature
decreases monotonously with the field, and the wave vector of the
instability starts to shift above  some critical value of magnetic field.
Depending on the
ratio between the spin and charge coupling constants and on the
direction of the  applied magnetic field, the wave vector shift is either
parallel ($CDW_x$  order) or
perpendicular ($CDW_y$  order) to the most conducting direction.  The
$CDW_x$
order
is a field  dependent  linear combination of the charge and spin density
waves and is
sensible only to the Pauli effect. The wave vector
shift in $CDW_y$ depends on the interchain coupling, but the critical
temperature
does not. This order   is affected by the confinement of the electronic
orbits.
By increasing the relative strength of the orbital
 effect with respect
to the Pauli effect, one can destroy the $CDW_y$, establishing
either a $CDW_x$, or a $CDW_0$ (corresponding to perfect nesting wave vector).
We also  show that
by increasing the imperfect nesting parameter, one  passes from the
regime where the
critical temperature decreases with the field to the regime where it is
initially enhanced by the orbital effect and eventually suppressed by the Pauli
effect. For a bad nesting, the quantized phases of the field-induced CDW
appear. }

\newpage

\section{Introduction}

The open and almost flat Fermi surface that characterizes the
quasi-one-dimensional  (Q1D) electronic systems gives rise to the formation
of charge (or spin)
density waves \cite{Rev1,Rev2,JerS}. Moreover, the external magnetic field
couples to the spin (via Pauli term) and to the
orbits (via Peierls substitution in the Hamiltonian) of the electrons. This
coupling affects  the properties
related to density wave ($DW$) formation  like the order parameter, the
critical
temperature and the wave vector of instability. The scale for  the Pauli impact
in the momentum space is the wave  number $q_P=\mu_BH/v_F$, while  the orbital
effect  enters  through the inverse magnetic length $q_o=ebH\cos \theta$,
where $\theta$ is the inclination  of the magnetic field {\bf H} from the
transverse $c$-direction in the $(b,c)$ plane (a plane perpendicular to
the chains). The ratio of these two characteristic wave  numbers
$\eta \equiv q_o/q_P=ebv_F\cos \theta /\mu _B$
is of the order of unity in real materials. It will play an important role in
the phase diagram for the $CDW$ in a magnetic field.

The Pauli term introduces a finite coupling between  the $CDW$ and the
component of the
$SDW$ parallel to {\bf H}, and may lead to a finite, magnetic
field dependent, shift in the wave vector of instability \cite{bz}.
It is therefore necessary to treat $CDW$ and $SDW$ together. A simple
relevant model is the extended Hubbard or $(g_1,g_2)$
model \cite{Rev1,JerS},  with coupling constants $U_s=g_2/2$ and
$U_c=(2g_1-g_2)/2$  for the SDW and CDW respectively.
Since the Pauli term mixes  the $CDW$ with the $SDW$, the ratio
$\nu \equiv -U_s/U_c$
will be the second relevant parameter for the $CDW$ phase diagram.

The Pauli term breaks the rotational symmetry of the complex vectorial
$SDW$ order parameter, constraining its direction perpendicularly to magnetic
field.  With this constraint taken into account, the  $SDW$
phase diagram depends only on the orbital coupling, provided that the system is
perfectly magnetically isotropic in the absence of magnetic field. However,
the fluctuations of the component of $SDW$ parallel to {\bf H} around its zero
value remain affected by both, Pauli and orbital coupling. Moreover the Pauli
term introduces a finite coupling between these fluctuations and the
noncritical  CDW fluctuations.

 The influence of a magnetic field on the $CDW$ systems is even richer, because
both Pauli and orbital  effects can affect the $CDW$ ordering. This fact is
of direct
experimental interest,  since e. g. the critical temperature can easily be
measured.
Furthermore, there  is a finite magnetic field at which the wave vector of
ordering starts to vary with the  magnetic field. The description of these
 features, together with the interesting
$CDW-SDW$ mixing, is the main objective of the present detailed analysis.

The various aspects of the interaction between the
electrons in Q1D systems and the external magnetic field have been already
subjects of  numerous analyses. The  quadratic  decrease of the mean-field
critical temperature in one-dimensional  $CDW$ systems due to the Zeeman
splitting
was  proposed theoretically \cite{DF}, and found experimentally in
the  organic compound
$TTF-TCNQ$ \cite{Tiedje}.  The recent very
precise measurements in $Per_2[Au(mnt)_2]$ \cite{Port} show
the  decrease of $T_c$ which differs considerably from the
theoretical value \cite{DF}. The effect of the Pauli
coupling on the $CDW$ order parameter can be formulated as a breaking of
degeneracy of two
density waves,  those with parallel and antiparallel spin with respect
to {\bf H}, each component being a $CDW$-$SDW$ hybrid.
This is reminiscent of the treatment of two coexisting $CDW$s with
overlapping electronic bands \cite{Braz,Nog,BB}.
The coupling of two $CDW$s with  different wave vectors may stabilize
a soliton lattice in the relative phase of two waves \cite{BB}.

On the other side, the orbital coupling alone leads to an increase of the
critical temperature for $CDW$s \cite{Balseiro,mon,bjma1}. Such an increase was
observed in e. g. NbSe$_3$ \cite{Coleman}.
The aim of the present work is to introduce both, orbital and Pauli couplings,
into the RPA calculation of the $DW$ matrix susceptibility, and to
determine some mean-field properties, in particular the phase diagram for
$CDW$ systems in a magnetic field.

In Sec.2  we  derive the RPA results  for  $DW$ response functions in the
form of a 4x4 matrix. In Sec.3  we analyse in detail the phase diagram for
$CDW$s
in the case of a perfectly nested Fermi surface. In particular we  consider
the influence of  the parameters $\nu$, $\eta$ and of the
interchain hopping $t_b$ on the critical temperature, the wave
vector of the instability, and the CDW-SDW coupling.  We also shortly
discuss
the effects of the
imperfect nesting on the  critical temperature as a function of magnetic
field.
The concluding remarks are given in Sec.4.

\newpage

\section { Model}
\mbox{}

Quasi-one-dimensional electrons in an  external magnetic field are usually
modelled  by the anisotropic two-dimensional Hamiltonian
\begin{equation}
\label{m1}
{\bf H}_o  = \frac{b}{2\pi}\int \; dq_y\; \int \; dx \; \Psi
^{\dagger}(x,\; q_y)\left[ H_{1D} + H_{Q1D,orb} + H_{Pauli}\right]\Psi
(x,\;  q_y)
\end{equation}
with
\begin{eqnarray}
\label{m2}
H_{1D} 		& = & iv_F\rho _3\partial _x		 \hspace{110mm}
(2a)\nonumber \\
H_{Q1D,orb}	& = & 2t_b\rho _3\sin (q_yb
-q_ox)+2t'_b\cos
2(q_yb -q_ox) \hspace{50mm}(2b) \hspace{700mm}\\
H_{Pauli}  	& = & -\sigma _3 \mu _B H 		\hspace{108mm}
(2c)\nonumber
\end{eqnarray}
Here $\Psi ^{\dagger}$ and $\Psi$ are four-component fermion fields,
$$
\Psi ^{\dagger}=\left( \Psi ^{\dagger}_{\uparrow +}\; ,
\Psi ^{\dagger}_{\uparrow -}\; ,
\Psi ^{\dagger}_{\downarrow +}\; ,\Psi ^{\dagger}_{\downarrow -}\right)
\nonumber
$$
where the indices $\uparrow ,\downarrow$ span the spin space  and
$\sigma _i$ are corresponding Pauli matrices. Indices $+(-)$ denote the
right (left) Fermi surface with the states defined with respect to
$\pm{\bf Q}/2 $, where ${ \bf Q} = (2k_F, \pi/b)$ is the wave vector of perfect
nesting realized for $t'_b = 0$, and
$\rho_i$'s are the Pauli matrices in that space. The chains
lie in the $xy$ plane and are parallel  to the $x$ axis.  $b$ is the latice
constant in the $y$ direction. The longitudinal electronic
dispersion given by $H_{1D}$ is linearized in the vicinity of the Fermi wave
numbers $\pm k_F$ , with $v_F$ being the longitudinal Fermi velocity.  $t_b$
is the hopping integral between nearest neighboring chains and $t_b'$
parametrizes the imperfect nesting. The spin space is
chosen to have  the third component parallel to {\bf H}.

Let us now introduce the relevant interaction part of the Hamiltonian. Since

the further
considerations are limited to the $2k_F$ RPA response, it is sufficient to
keep only
the contributions with bilinearly coupled electron-hole operators for spin
and charge density waves. They are given by
\begin{eqnarray}
\label{m3}
  {\bf H}_{int} & = & \int dx\: \sum _{{\bf R}_{\perp }}\;
[{-U_s{\bf M}}^{\dagger }
  ({\bf R})\cdot {\bf M}({\bf R}) +U_cM_4^{\dagger }({\bf R}) M_4({\bf R})]  .
\end{eqnarray}
The two-fermion operators in eq.(\ref{m3}) are defined by
\begin{equation}
\label{m4}
M_i=\Psi ^{\dagger}\rho _+\sigma _i\Psi \; \hspace{10mm}i=1,2,3,4
\end{equation}
where $\sigma _4\equiv I$. The first three components $(i=1,2,3)$
define the complex $SDW$ vector amplitude {\bf M}, while the fourth
component $M_4$  is the complex $CDW$
scalar amplitude. The $SDW$ and $CDW$ coupling constants in eq.(\ref{m3}) are
related to the usual backward ($g_1$) and forward ($g_2$) electron-electron
coupling constants by $U_s\equiv g_2/2$ and $U_c\equiv (2g_1-g_2)/2$.
We shall specify later the range of these constants for the most interesting
physical cases relevant for our analysis.

The mean field (MF) critical temperature for the spin or charge density wave
is defined as the temperature at which the corresponding RPA susceptibility
diverges.
In our case the Pauli term introduces a finite coupling between the component
of $SDW$ parallel to the magnetic field ($M_3$) and the $CDW$ ($M_4$).
This coupling is appropriately treated by introducing the
$DW$ susceptibility matrix, with  the elements defined as retarded correlators
\begin{equation}
\label{m5}
\chi _{ij}({\bf q},t-t') \equiv \langle M_iM_j^{\dagger} \rangle =
-\theta (t-t')\langle [M_i({\bf q},t),M_j^{\dagger}({\bf q},t')]\rangle
\;, \hspace{10mm}i,j=1,...,4\; ,
\end{equation}
where {\bf q} is the deviation of the wave vector from $(2k_F,\pi /b)$.
The RPA result for this matrix is \cite{bz}
\begin{equation}
\label{m6}
[\chi _{ij}({\bf q},\omega)]= \left(
\begin{array}{cccc}
\frac{\chi _o({\bf q},\omega)}{f^{\perp}} &                     0
               &                 0                  &           0 \\
              0                               &
\frac{\chi _o({\bf q},\omega)}{f^{\perp}} & 0 & 0 \\
                0                             &                          0
                 &
\frac{\chi _g(\sqrt{1+\delta ^2} +U_c\chi _g)}{f^{\parallel}}
     & \chi _g\delta /f^{\parallel} \\
                0                             &                          0
                 & \chi _g\delta /f^{\parallel}
      & \frac{\chi _g(\sqrt{1+\delta ^2} - U_s\chi _g)}{f^{\parallel}}
\end{array}
\right)\; ,
\end{equation}
with
\begin{eqnarray}
\label{m7}
  \chi _g   & \equiv & \sqrt{\chi _{\uparrow}({\bf q},\omega)\chi
_{\downarrow}({\bf q},\omega)}  \hspace{87mm}(7a)\nonumber \\
\chi _{\uparrow ,\downarrow}({\bf q},\omega _n)
	    & \equiv & \chi
_o(q_x\pm 2q_p,q_y,\omega _n) \hspace{87mm}(7b)\nonumber \\
\delta      & \equiv & [\chi _{\uparrow}({\bf q},\omega)-\chi _{\downarrow}
({\bf
q},\omega)]/2\chi _g \hspace{75mm}(7c)\hspace{75mm} \\
f^{\parallel}
            & \equiv & 1+(U_c-U_s)\chi _g \sqrt{1+\delta ^2}-U_cU_s \chi
_g^2 \hspace{59mm}(7d)\nonumber\\
f^{\perp}   & \equiv & 1- U_s \chi _o({\bf q},\omega)\; . \hspace{92mm}(7e)
\nonumber
\end{eqnarray}
 $\chi _o({\bf q},\omega)$ is the susceptibility  which includes
 orbital contributions of a magnetic field,
\begin{equation}
\label{m8}
\chi _o({\bf q},\omega _n)=\sum _{l=-\infty}^{\infty}P(q_x-lq_o,\omega _n)
I_{l}^2(q_y),
\end{equation}
where $P(k-lq_o,\omega _n)$ is the one-dimensional bubble. The coefficients
$I_{l}(q_y)$ bring in the orbital quantization due to the finite transverse
dispersion (2b) \cite{mon,bjma},
\begin{equation}
\label{m9}
I_{l}(q_y)= \sum _{l'} J_{l-2l'}\left( \frac{4t_b}{v_Fq_o}
\sin {\frac{q_yb}{2}}\right)J_{l'}\left( \frac{2t^{'}_b}{v_Fq_o}\cos q_yb
\right) ,
\end{equation}
where $J_l$ are Bessel functions.

Even without further diagonalization of the matrix (\ref{m6}), it
is evident that the critical temperatures for the condensation of density
waves
follow from the conditions
\begin{equation}
\label{m10}
f^{\perp}({\bf q}_{\perp},T_c^{\perp})  =  0
\end{equation}
and
\begin{equation}
\label{m11}
f^{\parallel}({\bf q}_{\parallel},T_c^{\parallel})  =  0 ,
\end{equation}
where all functions have to be taken in the static ($\omega=0$) limit.
$T_c^{\perp}$ is the critical temperature for the SDW with the orientation
of the spin perpendicular to {\bf H}, i.e. for the degenerate components
$M_1$ and $M_2$.  $T_c^{\parallel}$ is the critical temperature for the
hybrid of the $CDW$ and the $SDW$ with the spin parallel to  {\bf H},
i.e. of
the
coupled block $(M_3,M_4)$ in the matrix $[\chi _{ij}({\bf q},\omega=0)]$.
The corresponding wave vectors ${\bf q}_{\perp}$ and ${\bf q}_{\parallel}$
of the ordering are those which maximize the
respective critical temperatures $T_c^{\perp}$ and  $T_c^{\parallel}$. The
true critical
temperature of the $DW$ instability is equal to $max\{T_c^{\perp},
T_c^{\parallel}\} $.

Having in mind real systems,
it is appropriate to distinguish the most important situations realized for
two characteristic
interaction schemes. In the case of repulsive interactions
$(U_s>0\; ,\; U_c>0)$, usually
analysed in terms of the Hubbard model ($U_s = U_c > 0$),  the stable
ordering following from
(\ref{m7}) is the $SDW$ one,  determined by the condition (\ref{m10}). In other
words, as far as the system possesses the internal magnetic isotropy,
there is no effect
of Pauli coupling on the ordering. Its spin is oriented perpendicularly to
${\bf H}$, while the
wave vector is given by ${\bf q}_{\perp} = 0$ in the case of the good nesting
($t_b'\ll T_c$),
and may pass through the well-known cascade of phase transitions due to the
orbital
effects when the deviation from the good nesting is large enough ($t_b'\geq
T_c$) \cite{gor,FISDW,vcm,yam}.

In the case of  predominant electron-phonon interaction
($U_c < 0, |U_c| >U_s \geq 0$)
the system prefers the $CDW$ ordering.
As it is obvious from  eq.(\ref{m6}), the off-diagonal matrix elements vanish
in the absence of a magnetic field. The instability condition (\ref{m11}) then
reduces to $1 + U_c\chi_0 = 0$, and the ordering involves only the $CDW$
component $M_4$.
For finite magnetic fields the relation (\ref{m11}) contains
contributions originating from both, orbital and Pauli terms in the
Hamiltonian
(\ref{m1},\ref{m3}). The former enters through the bubble susceptibility
(\ref{m8}), while the latter introduces the $CDW-SDW$  (i.e. $M_4$ - $M_3$)
hybridization measured through the parameter $\delta$. As it is seen from
eq.(\ref{m7}),
$\delta$ is finite if $q_x \neq 0$. More explicitly, after diagonalizing the
$M_3 - M_4$
block of the matrix (\ref{m6})  the normal components of the "vector"
(\ref{m4}) read
\begin{eqnarray}
\label{m12}
M_- & = & \frac{1}{N}\left[ \delta M_3+ \Delta M_4\right] \nonumber \\
M_+ & = & \frac{1}{N}\left[ \delta M_4- \Delta M_3\right] \; ,
\end{eqnarray}
with
\begin{displaymath}
N   \equiv \sqrt{\delta ^2+\Delta  ^2}
\end{displaymath}
and
\begin{displaymath}
\Delta \equiv \frac{1-\nu }{2}
U\chi _g+ \sqrt{\left( \frac{1-\nu }{2} U\chi _g\right) ^2+\delta ^2},
\end{displaymath}
while the corresponding diagonal susceptibilities are
\begin{equation}
\label{m13}
\chi _{\pm}^{-1}=\chi _g^{-1}\left[\sqrt{1+\delta ^2}-
\frac{1 + \nu}{2}U\chi _g \pm \sqrt{\left( \frac{1 - \nu}{2}U\chi
_g\right) ^2+\delta ^2}\right]\; .
\end{equation}
In these equations we have defined
\begin{equation}
\label{m14}
U\equiv - U_c\;\;, \;\;\;\; \nu \equiv U_s/U = -U_s/U_c
\end{equation}
as a convenient parametrization of coupling constants for the problem of the

$CDW$ in the magnetic
field.  The value of $\nu$ depends on the interactions which participate
in the Hamiltonian (\ref{m1}, \ref{m3}).
The global phase diagram \cite{Rev1} at $H=0$ in $(\nu ,U)-$space   is shown
in Fig.1, where the
regime which we analyse is the upper half-plane (U$>$0).   The
superconducting (SC) instability which is present in this diagram is ignored
 in our RPA approach.
However,  since we are interested in the the effects of magnetic field,
this omission
can be justified   even in the case when for $H=0$ the singlet SC state (SS
in Fig. 1) is stable,
i.e. when $\nu <-1/3$.
Namely, the superconducting phase is suppressed by the orbital effect of a
magnetic field. The  critical  field  at
which the critical temperature
for the singlet SC state drops to zero is given by \cite{Dupuis}
\begin{equation} \label{HSC}
H_c^{SC}=\frac{16 \pi ^2T_{sc}^2(U,\nu )}{7\sqrt{2} \zeta (3)\mu _B \eta
t_b}\; ,
\end{equation}
where $T_{sc}$ is the critical temperature for the singlet superconducting
state  in the absence of magnetic field.
Considering $(g_1+g_2)$ as  the
corresponding effective coupling constant \cite{Rev1},
one easily finds that $T_{sc}$ is related to   the critical temperature for the
charge density wave at zero magnetic field, $T_c^o$, by
\begin{equation} \label{TSC}
T_{sc}=T_c^o\exp \left[ -\frac{\pi v_F}{U}\frac{1-3\nu}{1+3\nu}\right]\; .
\end{equation}
Equations  (\ref{HSC}, \ref{TSC})  give
 the estimation for
the magnetic field above which our RPA results are valid even in the regime
when the singlet superconductivity overwhelms the CDW.

It is useful for further discussion to mention here a
few characteristic
possibilities  regarding the value of the parameter $\nu$ in the $CDW$
(i.e. $U>0$) systems. Taking into acount only pure
backward electron-phonon interaction one has $\nu = 0$.
The inclusion of the presumably weaker
repulsive Coulomb interaction between electrons shifts $\nu$ to some
positive value. From
the other side, a pure Hubbard model with attractive on-site interaction
corresponds to
$\nu = -1$. Altogether,  $\nu$ covers a wide range of theoretically allowed
values, but it should be
noted that in the most frequent electron-phonon $CDW$ systems this range is
limited to $\nu \geq 0$.

Finally, it should be noted that the function $f^{\parallel}$ from the
matrix (\ref{m6}) can be
expressed in the factorized form,
$f^{\parallel} = \chi_g^{2}\chi_{-}^{-1}\chi_{+}^{-1}$.
Thus, for $U>0$ the condition (\ref{m11}) reduces to
\begin{equation}
\label{m15}
\chi_{-}^{-1}({\bf q},T_c) = 0,
\end{equation}
i.e. to the divergence of the susceptibility
$\langle M_{-}M_{-}^{\dagger} \rangle$. Indeed,
in the limit ${\bf H}\rightarrow 0$ the component  $M_{-}$ reduces to the
pure $CDW$
component $M_4$ and the divergence of $\chi_{-}$ coincides with the
condition
for the
$CDW$  instability, $1 + U_c\chi_0 = 0$.  Since the further discussion
involves only the ordering
with finite components $M_3$ and  $M_4$, we simplify the notation for
$T_c^{\parallel}$
and  ${\bf q}_{\parallel}$ in eq.(\ref{m15}) by skipping the index
$\parallel$.

\section {Discussion}
\mbox{}

For $t_b^{'} = 0$, the wave vector of $CDW$ ordering for ${\bf H} =0$ is
defined by the maximum
of the susceptibility $\chi_0({\bf q}, \omega_n = 0)$ (\ref{m8}) in
the limit $q_o \rightarrow 0$.
Of course it is located at ${\bf q} = 0$, i. e. at the wave vector of
perfect
nesting. The corresponding
critical temperature $T_c^0 = (2\gamma E_F/\pi )\exp (-\pi v_F/U)$  defines
the temperature
scale of the problem.

We want now to calculate the position of the minimum of $\chi _-^{-1}$
(eq.\ref{m13})
in the momentum space. The criterion of
local stability of the ordering with ${\bf q} = 0$ at finite ${\bf H}$  can
be derived from the
quadratic expansion of  $\chi _-^{-1}$  with respect to $q_x$ and $q_y$.
For $T=T_c(H)$ it suffices to expand the [..] bracket in eq.(\ref{m13}).
Noting that
there is no   bilinearly mixed term ($q_xq_y$), one gets
\begin{equation} \label{p1}
U^{-1}\chi_g\chi_-^{-1} \simeq U^{-1} - \chi _0 + a_xq_x^2 +
a_yq_y^2+b_yq_y^4+{\cal O}(q_x^2 q_y^2, q_x^4,..)\; ,
\end{equation}
with
\begin{equation}
\label{p2}
a_x = -\frac{1}{2}\frac{\partial ^2\chi _0}{\partial q_x^2} +
\left [\frac{1}{2\chi_0^2}(U^{-1} +\chi _0)
-\frac{1}{(1 - \nu )U^{2}\chi_0^3}\right] \left( \frac{\partial
\chi_0}{\partial q_x}\right)^2\; ,
\end{equation}
\begin{equation}
\label{p3}
a_y = -\frac {1}{2}\frac{\partial^2\chi_0}{\partial q_y^2},
\end{equation}
and
\begin{equation}
b_y=-\frac{1}{4!}\frac{\partial^4\chi_0}{\partial q_y^4}\; ,
\label{a4}
\end{equation}
with the values of $\chi_0\equiv\chi_0(q_x, q_y, \omega_n=0)$ and its
derivatives taken at
$q_x=2q_p,q_y=0$. For the later purposes we include one ($\sim q_y^4$) of
fourth
order terms in the expansion (\ref{p1}). Note that the expansion (\ref{p1})
is valid for $q_x<4\pi T_c^o/v_F, q_y<4\pi T_c^o/(t_bb)$.

The dependence of the critical temperature for the ordering at ${\bf q}=0$
on the magnetic field
follows from the equation
\begin{equation}
\label{p4}
U\chi_0 = 1.
\end{equation}
For small values of $H$ this expression reduces to the known result for the
suppression of the critical temperature due to the Pauli splitting of the
electron band \cite{DF},
\begin{equation}
\label{p5}
T_c=T_c^0 \left(1 -7\zeta (3) (\mu _B H/2\pi T_c^o)^2\right)\;  .
\end{equation}

The dependence of the coefficients $a_x, a_y$, and $b_y$ on the magnetic
field follows
straightforwardly from eqs.(\ref{m8}, \ref{m9}).  To this end we use the
relation
\begin{equation}
\label{p6}
P(q_x) = \frac{1}{\pi v_F}\left[\ln\frac{2\gamma E_F}{\pi T} -
Re\Psi(\frac{1}{2} +
\frac{i v_F q_x}{4\pi T}) + Re\Psi(\frac{1}{2})\right],
\end{equation}
where $\Psi$ denotes the digamma function, and expand the coefficients
(\ref{m9})
(with $t_{b}^{'}=0$) in terms of $q_y$ up to the quartic contribution.
Taking into account also eq.(\ref{p4})
one gets at $T=T_c(H)$
\begin{equation}
\label{p7}
a_x = \frac{1}{2\pi v_F}\left[\frac{\partial^2}{\partial q_x^2}
 Re\Psi - \frac{2\nu U}{(1-\nu )\pi v_F}
\left(\frac{\partial}{\partial q_x}Re\Psi\right)^2\right]_{q_x=2q_P}
\end{equation}
 with $\Psi \equiv \Psi(\frac{1}{2} +\frac{i v_F q_x}{4\pi T})$, and
\begin{equation}
\label{p8}
a_y  = \frac{1}{\pi v_F}\left( \frac{t_b}{2\pi T_c \eta h}\right)^2 b^2
\alpha _y,
\end{equation}

\begin{equation}
\label{a4psi}
b_y  = \frac{1}{\pi v_F} \left( \frac{t_b}{2\pi T_c \eta h}\right)^2 b^4
\left[\left( \frac{t_b}{2\pi T_c \eta h}\right)^2 \beta _y - \frac{1}{12}
\alpha _y \right]
\end{equation}

The coefficients $\alpha _y$ and $\beta _y$   in eq.(\ref{a4psi}) are given by
\begin{eqnarray}
\alpha _y&\equiv &  Re\Psi\left(\frac{1}{2}
+ih(1+\eta /2)\right) +Re\Psi\left(\frac{1}{2} +ih(1-\eta
/2)\right)\nonumber \\
& & -2Re\Psi\left(\frac{1}{2}
+ih\right)
\label{alfa2}
\end{eqnarray}
and
\begin{eqnarray}
\beta _y&\equiv &  \frac{1}{4}Re\Psi\left(\frac{1}{2}
+ih(1+\eta)\right) +\frac{1}{4}Re\Psi\left(\frac{1}{2}
+ih(1-\eta)\right) -\nonumber \\
 & & -\frac{1}{2}Re\Psi\left( \frac{1}{2}
+ih(1+\eta /2)\right) -\frac{1}{2}Re\Psi\left( \frac{1}{2} +ih(1-\eta
/2)\right)\nonumber \\
& &  +\frac{3}{2}Re\Psi\left( \frac{1}{2}
+ih\right) .
\label{alfa4}
\end{eqnarray}
respectively. Here we have introduced the dimensionless variable
$h\equiv\mu_B H/2 \pi T$. Note
 that the quantities $a_x$ and $a_y$ determine the
longitudinal  and transverse
 correlation lengths for $CDW$ fluctuations  ($\xi _x=U\sqrt{a_x}$ and
$\xi_y=U\sqrt{a_y}$) when
the temperature is close to $T_c$.

The wave vector of ordering stays at ${\bf q}= 0$ as far as the coefficients
$a_x(h)$ and $a_y(h)$ are positive, and starts to move in the longitudinal or
transverse direction when the former or later coefficient changes sign.
As it is seen from eqs.(\ref{p7},\ref{p8}), the function $a_x(h)/a_x(h=0)$
contains the interaction parameters  $U/v_F$ and $\nu$,
while the function $a_y(h)/a_y(h=0)$
depends only on the ratio $\eta=q_o/q_p=ebv_F\cos \theta /\mu _B$ which
measures the
relative impact of the orbital and Pauli coupling on the CDW.
Note that the parameter  $\eta$ can be  easily changed
by varying the angle $\theta$ between the direction of magnetic field and
the $c$ axis.
Since the reasons for possible deviations of stable components $q_x$ and
$q_y$ from zero
are essentially different, it is appropriate to consider each case separately.

{\em The longitudinal component of the CDW wave vector.}
The coefficient $a_x$  changes  its sign at
the critical field $h_{cx}\equiv\mu_BH_{cx}/2\pi T$  shown in
Fig.2.  For small values of  $\nu$ this dependence is given by
\begin{equation}
\label{p9}
h_{cx}\approx h_c^o\left(1-2.47 \nu \frac{U}{\pi v_F}\right)
\end {equation}
with  $h_c^o\equiv \mu _BH_c^o/(2\pi T) = 0.304$. As it is seen in Fig.2,
all curves $h_{cx}(\nu)$
pass through two common points, given by $\nu =0, h_{cx}=0.304$ and
$\nu = 1, h_{cx}=0$. At these points,  $h_{cx}$ does not depend on $U$.
The first point $(\nu =0)$ corresponds to the CDW ordering with the
SDW coupling equal to zero.
In the second case $(\nu =1)$ the interactions in the CDW and SDW channels
are of equal strengths
and opposite signs, i.e. we are at the  $CDW-SDW$ boundary. There, the
longitudinal splitting of the wave vector starts already at $h_{cx}=0$.
The change of sign of $a_x$ at $h_{cx}(\nu,U)$ causes a second order
transition from the  phase with {\bf q}=0  named $CDW_o$,
to a phase with {\bf q}=($q_x(h)$,0), named $CDW_x$. The dependence of
the wave vector on the magnetic field is the solution of the equation
$\partial \chi_-^{-1}/\partial q_x=0$, and can be written in the form
\begin{equation}
q_x=\frac {4 \pi T}{v_F} f_{\nu ,U}(h).
\label{tiltq1}
\end{equation}
The function $f_{\nu ,U}(h)$ is shown in Fig.3 for $U/\pi v_F=0.2$ and few
values of $\nu$.
In the limit $h>>1$ one has $q_x(h)\rightarrow 2\mu _B H/v_F$, as it was
already shown
previously in the case of repulsive Hubbard model ($\nu = -1$) \cite{bz}.

{\em The transverse component of the CDW wave vector.}
The critical field $h_{cy}$ at which a  finite transverse component of the
CDW vector develops
is shown in Fig.4. The line $h_{cy}(\eta)$ corresponds to the second order
transition
from $CDW_o$ to a phase $CDW_y$ with a transversely shifted wave vector.
At small values of $\eta$ the dependence $h_{cy}(\eta )$ is given  by
\begin{equation}
\label{p10}
h_{cy}\approx h_c^o\sqrt{1+0.088\eta ^2}.
\end {equation}
Here we use the approximative expression
 $Re \Psi \left(\frac{1}{2}+ix\right) \approx  \Psi \left(
\frac{1}{2}\right) + 8.414x^2(1+3.81x^2)^{-1}$,
valid for small values of the argument $x$.
The dependence of the wave vector component $q_y$ on $h$ for a fixed value
of $\eta$ can be represented by
\begin{equation}
\label{bq2exact}
q_y=\frac{2}{b}\arcsin \left[\frac{\pi T_c}{t_b}g_{\eta}\left( h\right)
\right],
\end{equation}
with the function $g_{\eta}\left(h\right)$ shown in Fig.5. Note that unlike
$h_{cy}$
the wave vector component $q_y$ depends on $t_b$.
 For small values of $h - h_{c y}$ the function $g_{\eta}$ reduces,
after using eq.(\ref{p1}), to
\begin{equation}
\label{fetaaprox}
g_{\eta}\left( h\right) \approx 2 \eta h \sqrt{-\frac{\alpha _y}{2\beta _y}},
\end{equation}
with $\alpha _y$ and $\beta _y$ given by eqs.(\ref{alfa2}) and (\ref{alfa4})
respectively.  On the other hand in the high field limit $h\gg h_{cy}$ and
for $\eta =0$ the function $g_{\eta =0}(h)$  is asymptotically given by
$g_{\eta =0}(h\gg 1)\rightarrow h +\kappa$, where $\kappa$ is of the
order $1/\pi$.  Note that the transverse  shift of the wave vector does
not depend on the interaction ($\nu$ or $U$).
It however depends on $\eta$, i. e. on the relative impact of
the Pauli and orbital effects. The  reason for  this
is in the fact that all interaction dependence enter with $\delta$
(see eq.\ref{m7}),
which is equal to zero if  $q_x=0$ and if the nesting is perfect.
Thus, only a phase
$CDW_x$ is affected by the  finiteness of the $SDW$ coupling constant
$U_s$ (i. e. $\nu$).

{\em The phase diagram for $h$ larger than $h_{cx}$ and/or $h_{cy}$.}
To provide some ideas on the variation of the wave vector of instability
at magnetic fields larger than critical values $h_{cx}$ and
$h_{cy}$, it is useful to consider the symmetry and the shape of functions
${\chi _0({\bf q}),\chi _g({\bf q})}$ and ${\chi _-^{-1} ({\bf q})}$ at
 strong magnetic fields ($h$ of  order 1).
Note firstly that all these functions are even in $q_x$ and $q_y$.
The function ${\chi _0^{-1}({\bf q})}$
at $T<<t_b$ has the line of local maxima given by
\cite{Gillesthese}
\begin{equation}
q_x=\pm \left[ \frac{4t_b}{v}\sin{\frac{q_yb}{2}}+\frac{1}{v}{\cal O}\left(
\frac{T}{2t_b}\right) \right].
\end{equation}
When the Pauli term is introduced, the maxima of
${\chi_{\uparrow}({\bf q})}$ will move to the left and those of
${\chi_{\downarrow}({\bf q})}$ to the right by ${2q_P}$
along the axis $q_x$. The lines of local maxima
of the  susceptibilities $\chi _o$, $\chi _{\uparrow}$ and
$\chi_{\downarrow}$  are
shown in Fig.6a. For $h$ large enough the function
${\chi _g({\bf q})=\sqrt{\chi_{\uparrow}\chi_{\downarrow}}}$,
together with the  function ${-\chi _-^{-1}({\bf q})}$, will have two pairs of
degenerate maxima in {\bf q}-space as candidates for absolute maxima.
These two pairs have approximate positions at  $(\pm 2q_P,0)$
and $(0,\pm \frac{2}{b}\arcsin (v_Fq_P/2t_b))$  (denoted as A,A' and B,B'
respectively in  Fig.6a), in accord with the asymptotic limits given by
eqs.(\ref{tiltq1},\ref{bq2exact}).
In Fig.6, we also show the function ${-\chi _-^{-1}({\bf q})}$ for three
characteristic choices of parameters $\nu$, $\eta$ and $U$, i.e.  when the
absolute maxima are at $(0,\pm q_y)$ (Fig.6b),  at $(\pm2q_P,0)$
(Fig.6c), and when the two pairs of maxima have the same value (Fig.6d).
As it was shown above, the phase transitions from the $CDW_o$ to
$CDW_x$ and $CDW_y$  (Figs. (6b) and (6c)) are of the second order.
The transition between the orderings $CDW_x$ and $CDW_y$, caused
by the competition of two maxima in ${-\chi _-^{-1}({\bf q})}$  (Fig.6d)
is of the first order since the wave vector  has a discontinuous jump between
points $(q_x, 0)$ and $(0, q_y)$ (i.e. between points A and B in Fig.6).

To complete the phase diagram it is necessary to calculate
the magnetic field dependence of the critical temperatures, defined as the
solutions of eq.(\ref{m5}) for ${\bf q}=(0,0)$, ${\bf q}=(q_x,0)$
and  ${\bf q}=(,q_y)$, and denoted by $T _0$, $T_x$ and
$T_{y}$ respectively, and to determine $max[T _0(H),T_x(H),T_{y}(H)]$.
The dependence of critical temperatures $T _0, T_x$ and $T_{y}$ on
H for few values of $\nu$ and $\eta$ and for $U/\pi v_F=0.2$ is shown in
Fig.7a.
The sections of lines $T_x(H)$ and $T_{y}(H)$ determine the critical
magnetic fields and the temperatures of the first order transitions. Note
that the present analysis is based on the Landau expansion
\begin{equation}
\label{Landau}
F=\int d^2q \;  \chi _-^{-1}({\bf q})[M_-({\bf q})M_-^*({\bf q})]+
{\cal O}(\{ M_-^4 \} ),
\end{equation}
which is restricted to the range of temperatures not far below
$max[T _0(H),T_x(H),T_{y}(H)]$.

Since the complete phase diagram depends on three parameters, $H, \nu$
and $\eta$ (with fixed  $U$),  it is appropriate to use
two planes, $(H,\nu )$ ($\eta$ being a parameter)
and $(H,\eta )$ ($\nu $ being a parameter), for its presentation,
as shown in Figs. (7b) and (7c)
respectively.  We stress a particularly interesting
situation $\nu \rightarrow 1^-$
for which the critical field $h_{cx}$ goes to zero, and three phases
($CDW_0, CDW_x$ and $SDW$) are present in the narrow range of parameter
$\nu$.  Note also the presence of the point in Figs.(7b) and (7c)
at which the $CDW_0$, $CDW_x$ and $CDW_y$ orders meet. The dependence
$\nu(\eta)$ which defines this tricritical  point is shown in Fig.(7d). The
corresponding magnetic field weakly varies with $\nu$ (i. e. $\eta$), as
is seen in Figs.(7b) and (7c). The line in Fig.(7d) thus divides the region
where the
wave vector shifts firstly in transversal direction
from the region in which only a longitudinal shift is possible.
Furthermore,  among the $CDW$ phases from Figs. (7b,c) only the phase
$CDW_x$ has
a finite fraction of the component $M_3$ [see eq.(\ref{m12})], and is thus a
$CDW-SDW$ hybrid. The ratio of components $M_3$ and $M_4$ follows from the
constraint $M_+=0$. At $T=T_{c x}$  it is given by
\begin{equation}
\label{p13}
M_3({\bf q})=\frac{\delta ({\bf q})}
{\sqrt{1+\delta ^2 ({\bf q})} -\nu U\chi _g({\bf q})}M_4({\bf q}),
\end{equation}
and shown in Fig.8  for few values of $\nu$. Note that
$|M_3/M_4|$ tends to 1 as one  approaches the $CDW_x-SDW$ transition.

{\em Influence of the imperfect nesting.}
Let us finally consider a case when the imperfect nesting is introduced
through a finite  effective next-nearest neighbor hopping $t_b'$, which can be
usually increased by e. g. applying a strong pressure on a CDW system. For
example
the relevant pressure scale in $NbSe_3$ is about 10 kBar\cite{mon,Nunez}.

At small values of the magnetic field, the the critical temperature
for the phase $CDW_0$  can be readily found from the eq.(\ref{p4}) yielding
\begin{equation}
\label{Tcimp}
T_c-T_c^o\approx -\left( \frac{\partial \chi _o}{\partial T} \right)
\left[ \frac{1}{2} \frac{\partial ^2 \chi _o}{\partial q_o^2}q_o^2 -4 a_x
q_p^2 \right] .
\end{equation}
The values of $\frac{\partial \chi _o}{\partial T}$ and $\frac{\partial ^2
\chi _o}{\partial q_o^2}$ as functions of $t_b'$ are given in
ref.\cite{mon}. For small $t_b'$, the coefficient $a_x$  is given by
\begin{equation}
\label{ax}
a_x\approx \frac{v_F}{32 \pi ^3 T_c^o}\left[ -\Psi ''(1/2)+\Psi
^{IV}(1/2)\left( \frac{t_b'}{2\pi T_c^o}\right )^2 \right ] ,
\end{equation}
where $\Psi ''\approx -16.83$
and $\Psi ^{IV}\approx -771.47$. As one sees
from the eq.(\ref{Tcimp}), the orbital and Pauli effects are in competition,
the former trying to enhance, and the latter  to suppress $T_c$.  For small
$t_b'/T_c^o(t_b'=0)$ the function $\frac{\partial ^2 \chi _o}{\partial
q_o^2}$ is proportional to ${t_b'}^2$. Moreover, the imperfect nesting
decreases the coefficient $a_x$. Altogether, the general trend of the small
$t_b'$ is to flatten the $H$ dependence of $T_c$.

For the sake of space,  we present the result for the critical temperature
wich follows
from the eq.(\ref{m11})  only for the case of attractive
Hubbard interaction (i.e. $U/\pi v_F=0.2$; $\nu =-1$)\cite{nu} and for $\eta
=2.5$. In this regime the orbital effects are strong enough,
which excludes the stabilization of the $CDW_y$ ordering when the nesting is
good.
The interplay between  two effects of a magnetic field  is  a main
characteristic of
the  phase diagram for imperfect nesting, given in Fig.9.
As the parameter $t_b'$ increases from zero the critical
temperature  only monotonously shifts to lower temperatures, still decreasing
with a magnetic field. In other words, our results for the perfect nesting
can be applied even to the systems with a moderate finite imperfect nesting,
i.e. when the critical temperature remains  far above the value of $t_b'$.
The orbital effects  enter manifestly into play
at rather large values  of $t_b'$, enhancing the critical temperature
initially, as it was  observed in $NbSe_3$ \cite{Coleman}. The eventual
supression of $T_c$ by Pauli term at high magnetic fields  will make these
diagrams
basically different from the mean-field one for the FISDW with the orbital
coupling only \cite{FISDW}, where no eventual supression of the $T_c$ is
present.
For a very bad nesting,  i. e.  for $t_b'$  comparable to $t_b'^*$ [where
$ t_b'^*\sim T_c^o(t_b'=0)$) is the imperferct nesting parameter at which
the  $CDW$
is destroyed  at zero field \cite{mon}], the cyclotron frequency becomes the
first
relevant energy scale, giving  the rise to a cascade-like
shape, associated with the quantized field induced CDW phases. Notice that
our approach
does not explain the strong field breakdown of the high field  phase in
$(TMTSF)_2Cl0_4$ \cite{Chaikin}, since the  Pauli term does not affect the
$SDW$.

\section{Conclusion}

The main result of the present work concerns the phase diagram of a CDW
system in an external magnetic field.
There are three physical parameters which characterize this diagram, namely
the ratio of the SDW and CDW coupling constants, the strength of the magnetic
field and its direction with respect to the most conducting plane $(x, y)$.
The
respective parameters are $\nu$, $h$ and $\eta$. We recall that $\eta$
also measures the relative impact of the orbital coupling with respect to
the Pauli coupling.

In the case of good nested Fermi surface   the wave vector of
the CDW has a general tendency to shift
from its zero field value $(2k_F,\pi /b)$ as the magnetic field increases
[see Figs.(7a-c)].
This shift starts continuously, and may occur either longitudinally or
transversally with respect to the chain
direction. The longitudinal shift is governed solely by the Pauli coupling,
with
the corresponding $CDW_x$ state being a hybrid of the pure $CDW$ and of the
$SDW$ component parallel to the magnetic field.
Both, the critical value of the magnetic field $h_{cx}$ at which $q_x$
starts to shift, and the relative weights  of the $CDW$ and the $SDW$,
depend on the ratio  $\nu$. Both, $q_x(h)$ and the CDW-SDW hybridization
increase with the magnetic field.
It is important to mention that $h_{cx}$, $q_x$ and the hybridization
ratio  do not depend on $t_b$ because all mean
field properties concerning a  longitudinal tilt of the  wave vector
are given by pure one-dimensional expressions.

The shift of the $CDW$ wave vector in the transverse direction is affected
by both orbital and Pauli  couplings. Contrary to the $CDW_x$, the $CDW_y$
is not a $CDW-SDW$ hybrid, and therefore is not influenced by the parameter
$\nu$.
It exists only when $t_b$ is finite, although the critical magnetic field
$h_{cy}$ does not depend on $t_b$.
 However,  $t_b$
influences the variation of $q_y$ at $h>h_{cy}$, as shown
by eq.(\ref{bq2exact}).
$q_y(h)$ decreases with $t_b$ and increases with  the
magnetic field. According to the general fact that the orbital effects
lower the dimensionality of the electronic motion \cite{gor}, the effect
of the increasing $\eta $ is to favor the $CDW_{x}$. After some
critical value  of $\eta$ (dependent on $\nu $), the orbital
impact reduces  the phase diagram to the pure one-dimensional one,
consisting  only of the $CDW_o$ and  $CDW_x$, as it is seen from Fig.7b.

At $\eta =0$ and for $\nu < 0$, the shift of the wave vector is at  first
directed perpendicularly, and jumps to the longitudinal direction at
some higher magnetic field. This jump between $CDW_y$ and $CDW_x$
is a first order transition. On the contrary, for $0 < \nu < 1$, the wave
vector  is shifted longitudinally for all magnetic fields higher than
the critical field $h_{cx}( \nu)$. Furthermore, $h_{cx}( \nu)$
tends to zero as $\nu$ approaches unity. The point $H = 0, \nu = 1$
is therefore tricritical, since $\nu > 1$ is the range of SDW stability.

The Pauli and orbital terms together cause a rather
complex magnetic field dependence of the critical temperature in
systems with a finite imperfect nesting.  This is
illustrated in Fig.9 in which $t_b'$ varies from zero to the range
above the critical value $t_b'^*$, at which the CDW ordering is
completely eliminated at zero magnetic field. A rich dependence  $T_c(H)$
contains the
suppression  by the Pauli term, enhancement  by the orbital effects and, for
large  values of $t_b'$,  a cascade-like shape characterizing the
field-induced DW. This
phase diagram is quite general and not limited to the value $\nu =-1$,
chosen in Fig.(9).

Our analysis for the perfect nesting case, showing a strong dependence of the
critical properties in magnetic field with the ratio $\nu$, could find
an appropriate experimental support e.g.  in the MX compounds.
The low-dimensional
nature of these materials corresponds to our model. From  our
analysis, a  particularly interesting  possibility is that the Coulomb and
electron-phonon forces  can be tuned in a predictable manner by external
pressure
\cite{MX0} or chemically \cite{MX}, allowing  to approach the phase
boundary between
CDW and SDW, corresponding to $\nu =1$.
As we approach  the boundary from the CDW side, the
critical field for the $CDW_o \rightarrow CDW_x$ transition
$h_{cx}\equiv H_{cx}/2\pi T$ will decrease rapidly toward zero,
regardless of the value of $T\approx T_c$. Even for large $T_c$,
by adjusting carefully $\nu$,   $H_{cx}$ can decrease to   experimentally
reachable values, being
extremely sensible to the variation of the parameter $\nu$.
We point out that a search
for a magnetic field induced phase transitions in a CDW phase with strong
SDW fluctuations (introduced by high pressure, for example) could confirm
our predictions.

In $NbSe_3$ a phase transition in the 59K CDW phase induced by
magnetic field was found \cite{MoncRich} by observing
that a threshold electric field for the collective CDW motion
is strongly reduced when magnetic field increases beyond the critical point.
The naive explanation that this is a simple $CDW_o\rightarrow
CDW_x$ one-dimension-like transition due to only Pauli term must
be taken with caution. Namely, the observed effect strongly depends on the
angle $\theta$, indicating that the orbital effects are also involved. This
might mean that the strong orbital contributions, provided by a badly nested
Fermi surface, affect the phase diagram. However, we believe that the
Pauli term has an important role in this transition, since it enables the
shift of
the wave vector from its  commensurable,  perfect nesting position. We remind
that pure orbital effects can affect the wave vector only if it is not at
the perfect nesting position [like  in e.g. $(TMTSF)_2ClO_4$ \cite{Chaikin}].
The fact that the nesting in $NbSe_3$ is quite bad can be deduced
from relatively strong pressure dependence of $T_c$ $(dT_c/dP\approx
-6.25 K/Kbr$)\cite{Nunez}. Indeed,  from the comparisson of a
very weak enhancement of $T_c$ with magnetic field \cite{Coleman} with
our results in Fig.9, it follows that the value of $t_b'$ should be rather
large.

Finally, our analysis of the imperfect nesting case can somewhat enlighten
the recent measurements \cite{Port} in the compound $Per_2[Au(mnt)_2]$,
where the suppression of the critical temperature proportional
to the square of the magnetic field was found,  but with a
coefficient smaller than that which follows after taking only the Pauli
coupling and a
perfectly nested Fermi surface \cite{DF}. From eq.(\ref{Tcimp}) and from
Fig.9 one can conclude  that  the
reason for the flattening of the suppression of $T_c$  is just the
finiteness of
$t_b'$. However, the situation is not  so simple.  At finite
values of   $t_b'$
the orbital effects come into play,  in contrast to the experimental results
which are
independent on the field direction.
If we just ignore the orbital effects, we get $t_b'\approx 7.4K$ as a
imperfect nesting parameter fitting the experimental curve. Finally, we
indicate that the
measurements of the critical properties in a  magnetic field,  and  with
pressure
large enough to almost or completely destroy the zero-field CDW,    could show
very strong, cascade-like enhancement of the $T_c$ for the quantized field
induced $CDW$ phases.

\bigskip

{\bf Acknowledgments}

We  are indebted to S. Brazovskii, N. Dupuis  and J.-P. Pouget
 for inspiring discussions,
and to M. Latkovi\'{c} for the help in some of the numerical analyses.

\newpage
\begin{center}
{\bf FIGURES}
\end{center}

FIG. 1. A phase diagram of  the one-dimensional system in $(\nu ,U)$ plane,
in the absence of a magnetic field.

FIG. 2. Scaled critical  magnetic field $h_{cx}\equiv \mu _BH_{cx}/(2 \pi T$)
as a function of
the parameter $\nu$ for few choices of the coupling constant: $U/\pi v_F
=$0; 0.2; 0.4; 0.6.
 Note that one has to insert $T_c(H)$ [and not $T_c(H=0)$] into the
defining expression for $h_{cx}$ in order to get a phase diagram with
H dependence.

FIG. 3. The function $f_{\nu, U}(h)$, determining  the dependence of the
 longitudinal shift of the wave vector on the magnetic field
(see eq.\ref{tiltq1}), for $U/\pi v_F=0.2$ and
 $\nu =-1$ (A), -0.5 (B), 0. (C), 0.25 (D), 0.5 (E), 0.75 (F), 0.99 (G).

FIG. 4. Scaled critical  magnetic field $h_{cy}\equiv \mu _BH_{cy}/(2 \pi
T$)  as a function of the parameter $\eta$.

FIG. 5. A function $g_{\eta}(h)$, determining  the dependence of the
transverse shift of the wave vector on the magnetic field (see
eq.\ref{bq2exact}), for $\eta =0$ (A), 0.5 (B), 1 (C), 1.5 (D), 2 (E).
The inset shows
the large $x$ behavior for a case with no orbital effects ($\eta =0$).

FIG. 6. a) The lines of local maxima in {\bf q}-space of the
susceptibility $\chi _o(q_x, q_y)$ (full line) without magnetic field, of
$\chi _{\uparrow}\equiv \chi _o(q_x+2q_P,q_y)$ (dot-dashed line) and of
$\chi _{\downarrow}\equiv \chi _o(q_x-2q_P,q_y)$ (dashed line).
(A,A')
and (B,B') are the two pairs of degenerate maxima of $\chi _g(q_x, q_y)$.
The figures b), c) and
d) show the function $-\chi_-^{-1}({\bf q})$ at $T=0.42T_c^o$ and $\mu
_BH=1.14T_c^o$ respectively for three cases: $\nu =-1$ ($CDW_{y}$ is stable,
provided that the maxima are in the points B,B'),
$\nu =-0.1$ ($CDW_x$ is stable; the points A,A' are dominant) and
$\nu=-0.33$ (the first order
critical
point between $CDW_{y}$ and $CDW_x$, since all four points, A,A',B and B'
are of the equal height).

FIG. 7. a) The critical temperatures $T_0$, $T_x$ and $T_y$ for the CDW
instabilities  with the wave vectors {\bf q}= 0 (full line), {\bf
q}=($q_x,0$) (dotted line) and {\bf q} = ($0, q_y$) (dot-dashed line)
respectively. For all curves $U/\pi v_F=0.2$.

b) The phase diagram in $(\nu ,\mu BH/(2\pi T_c^o))$ plane
for $\eta = 0$.
The changes of the diagram with finite $\eta $ (here, $\eta = 1$) are shown
by dashed lines.

c) The phase diagram in $(\eta ,\mu BH/(2\pi T_c^o))$
plane for $\nu =-1$.  The changes of the diagram when $\nu =-0.5$ are dashed.

d) The curve in the $(\nu, \eta)$
plane which defines the tricritical point in Figs.(b) and (c).

FIG. 8. The relative weight of the $SDW$ and $CDW$ components in the hybrid
phase $CDW_x$ as a function of magnetic field, for $U/\pi
v_F=0.2$ and for $\nu =-1$ (A), 0 (B), 0.99 (C).

FIG. 9.  The critical temperature {\em vs} magnetic field for a series of
values of $t_b'/t_b'^*$ and for $U/\pi v_F = 0.2$ and $\nu = -1$.

\end{sloppypar}

\begin{thebibliography}{99}
\bibitem{nu} $\nu =0$ actually corresponds to a stable SC phase. However,
the form of the phase diagram is expected to be almost the same for all
negative values of $\nu$.
\end{thebibliography}
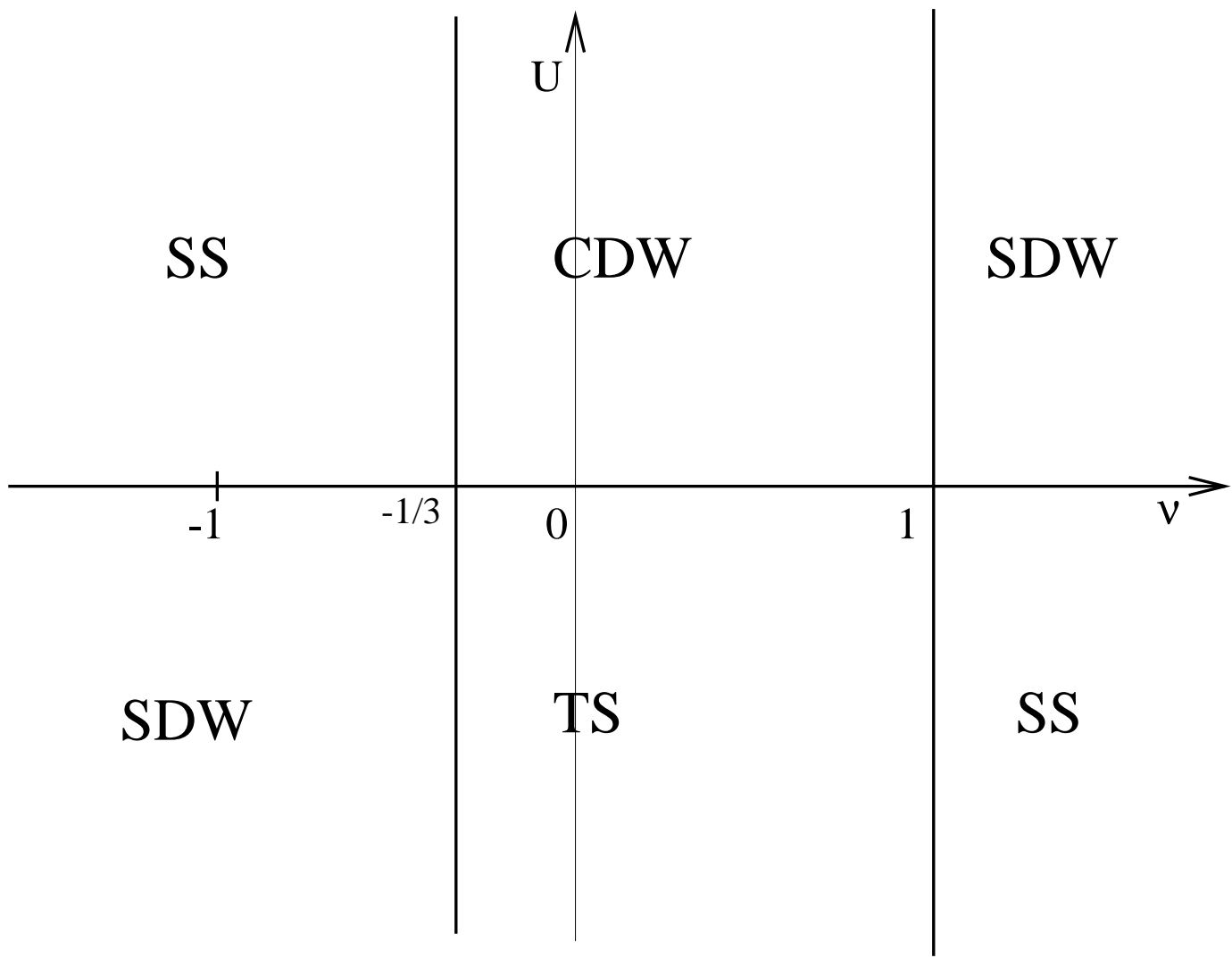
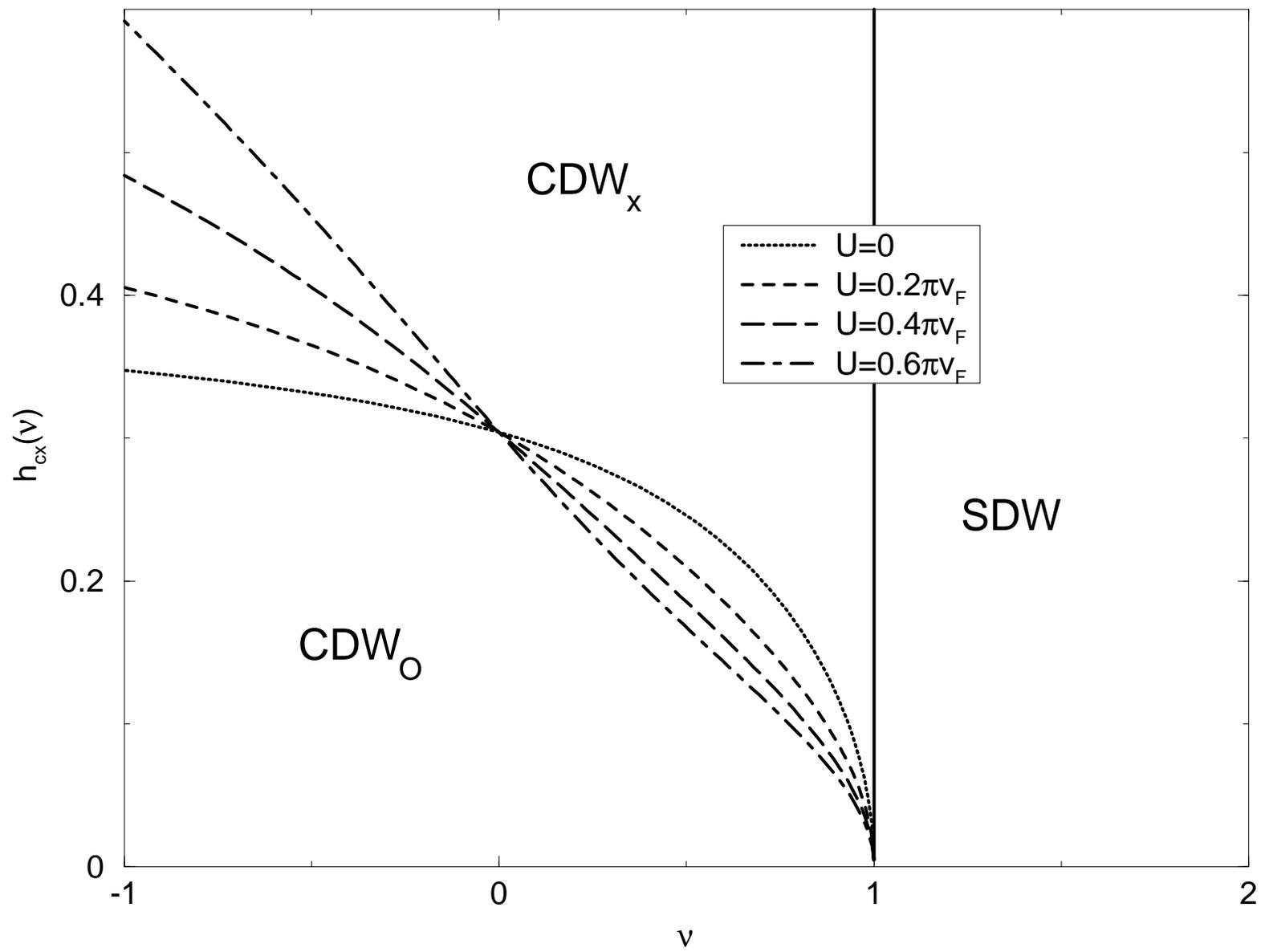
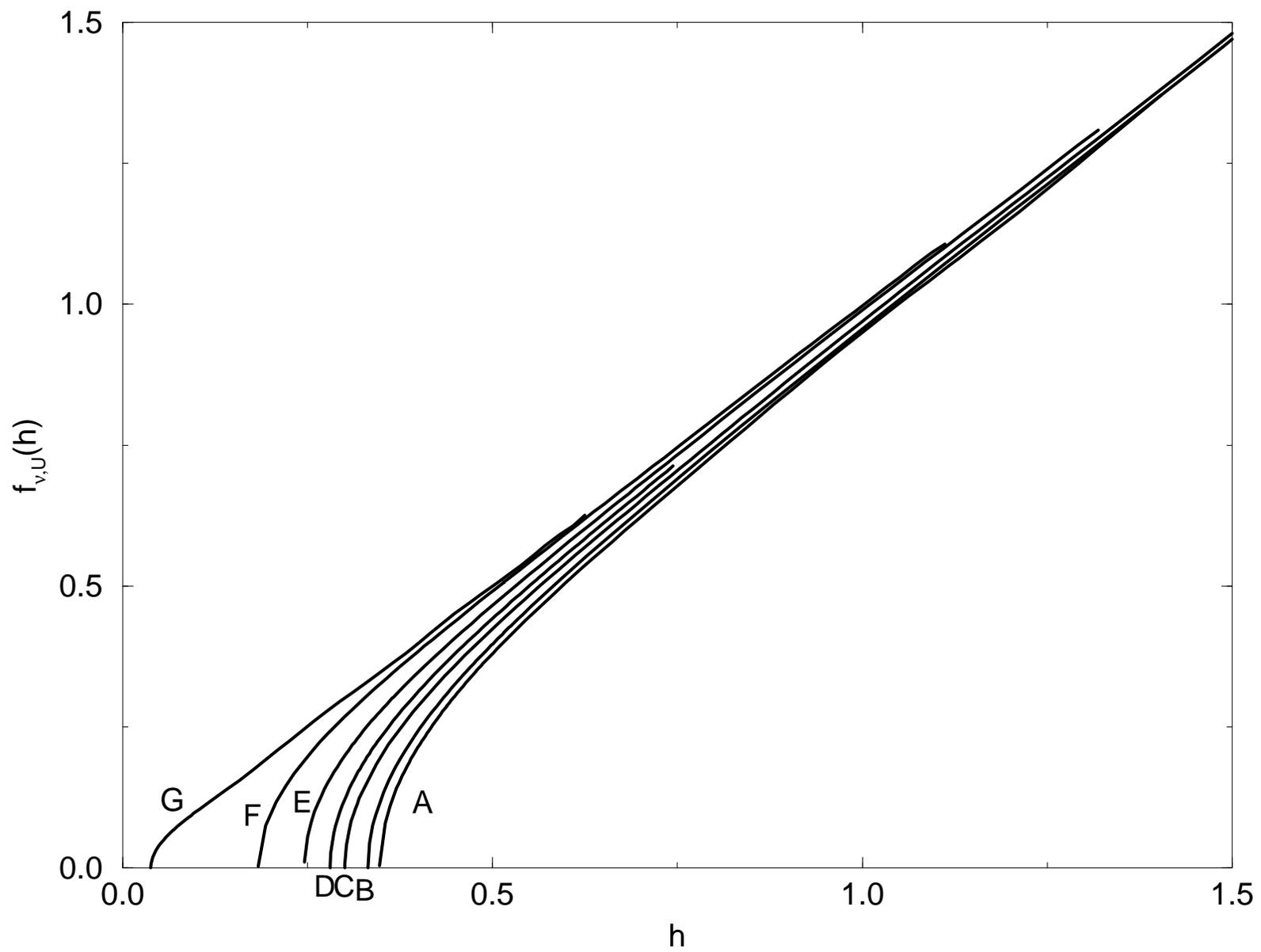
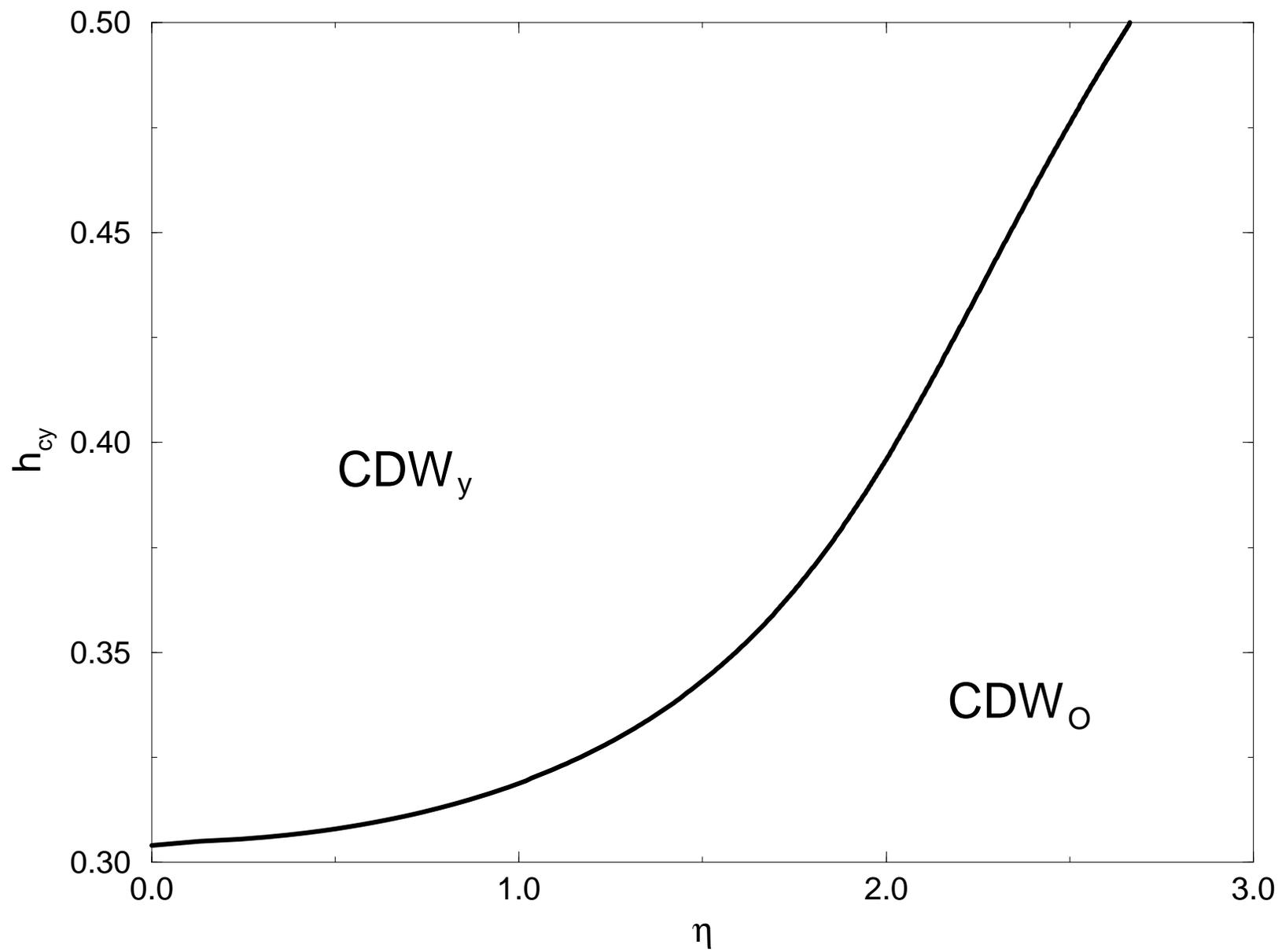
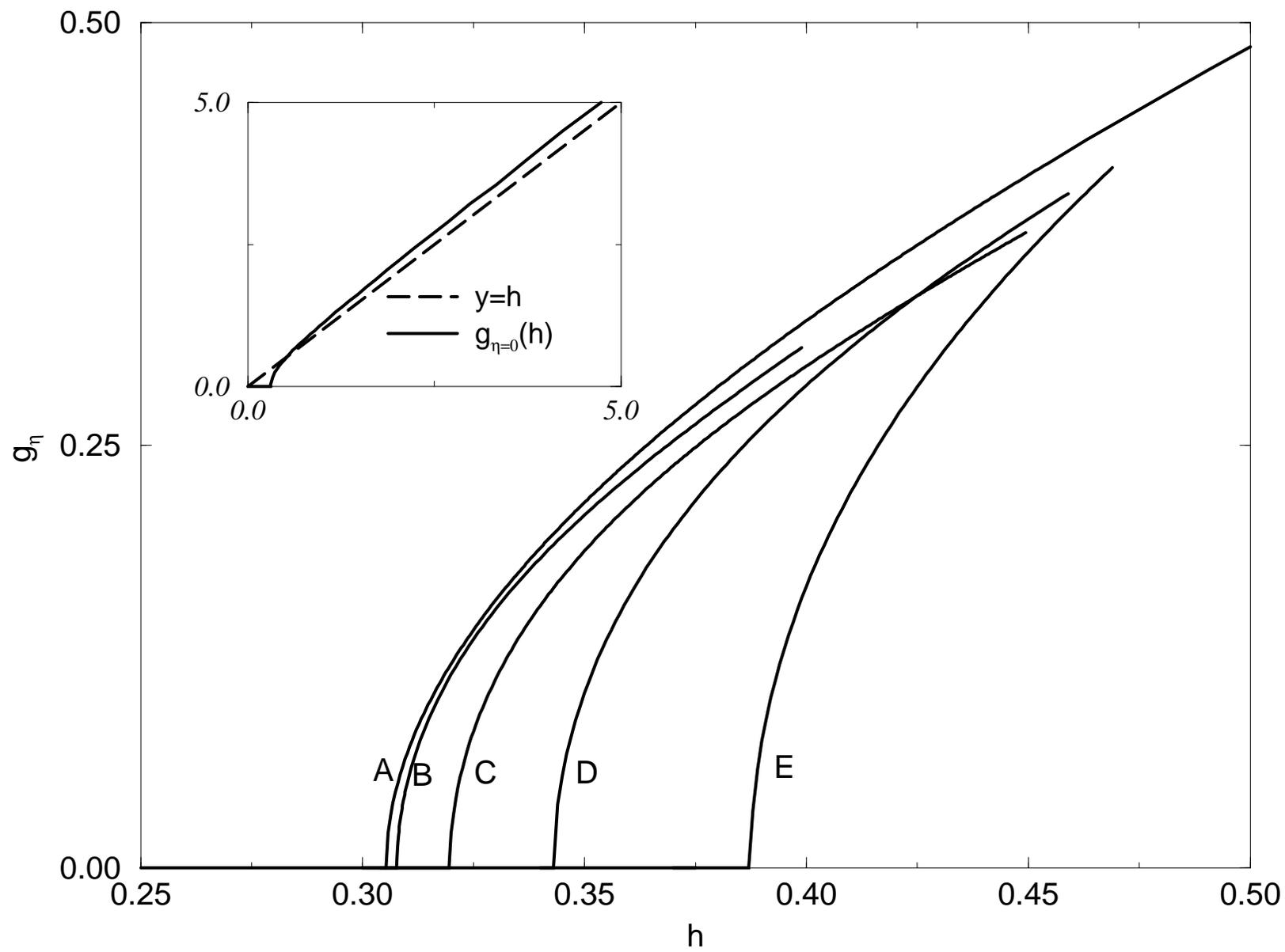
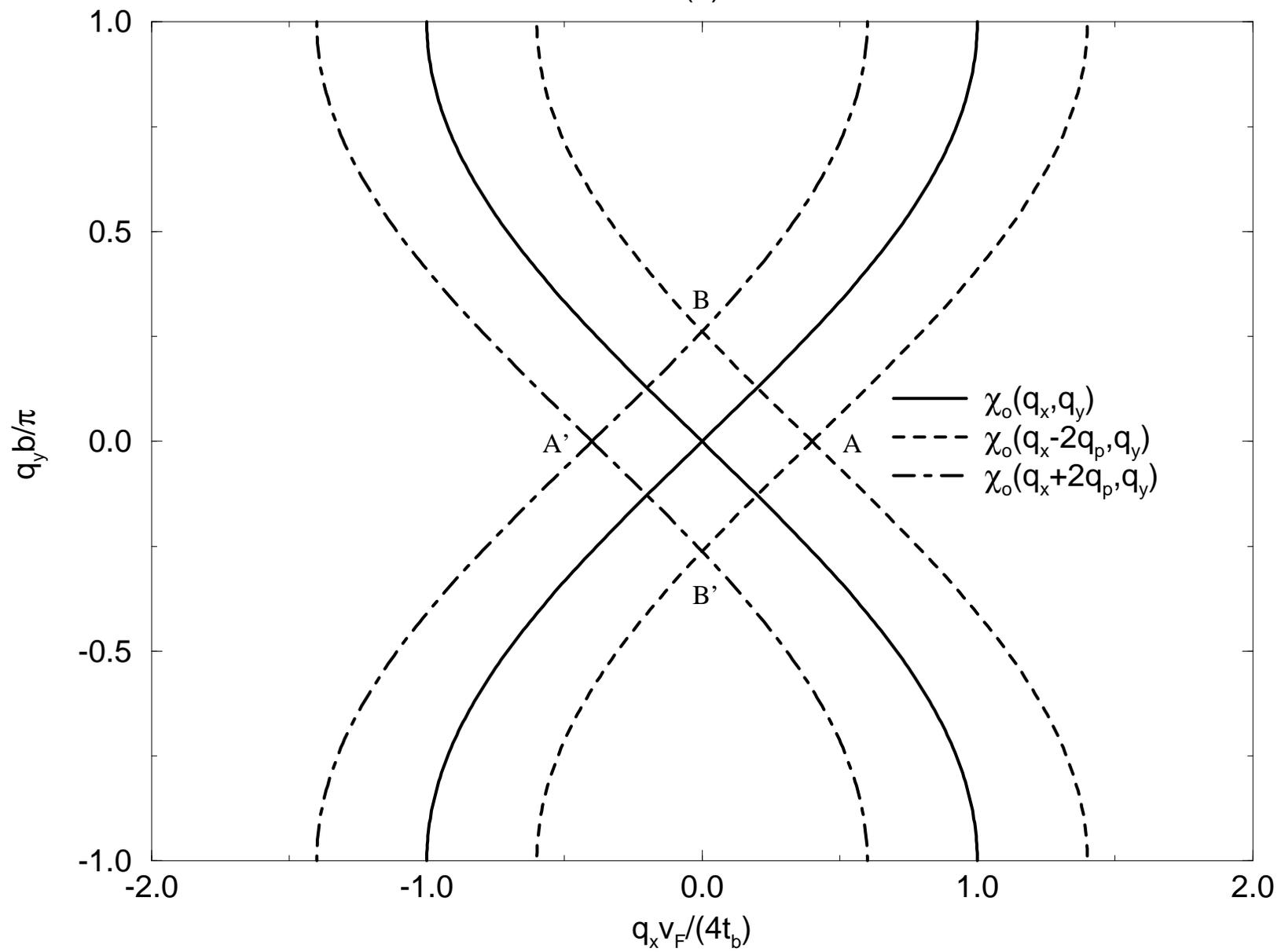
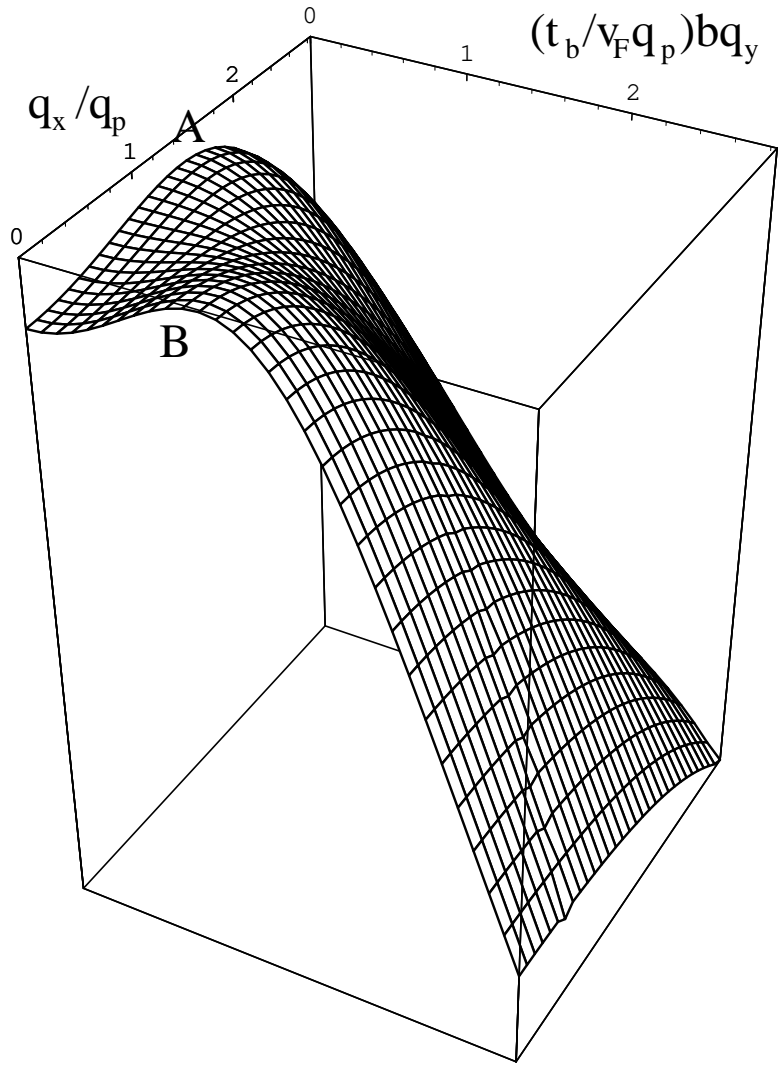
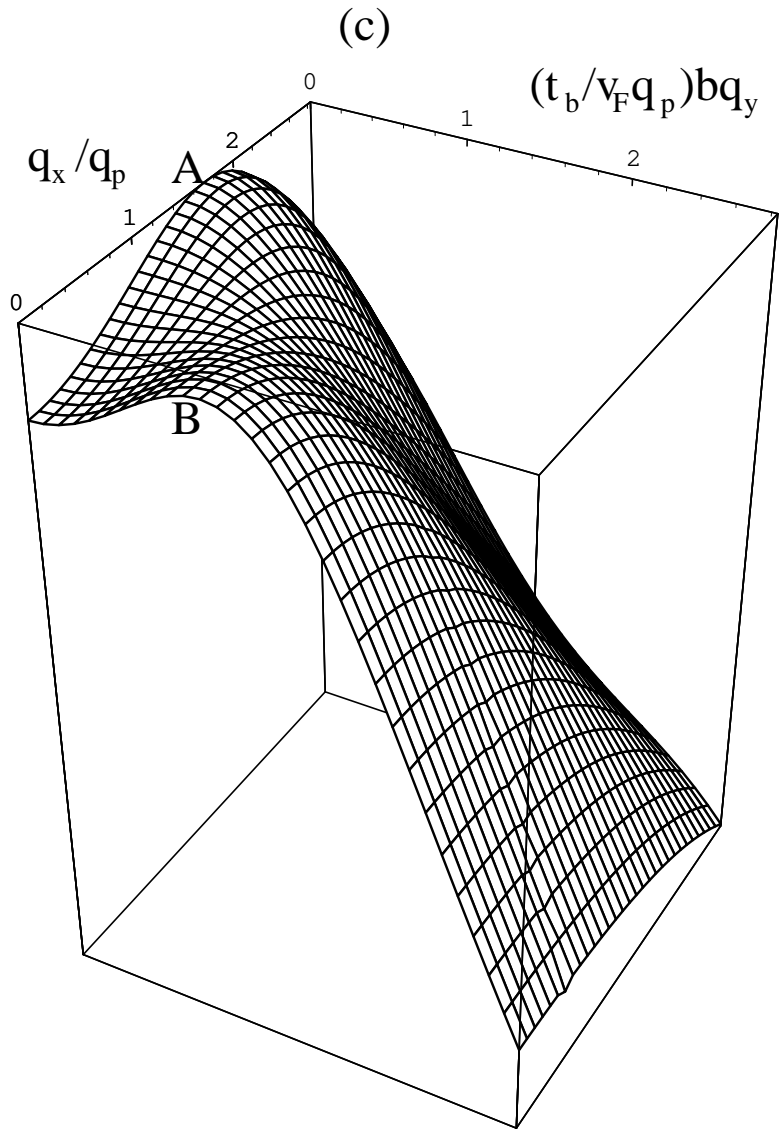
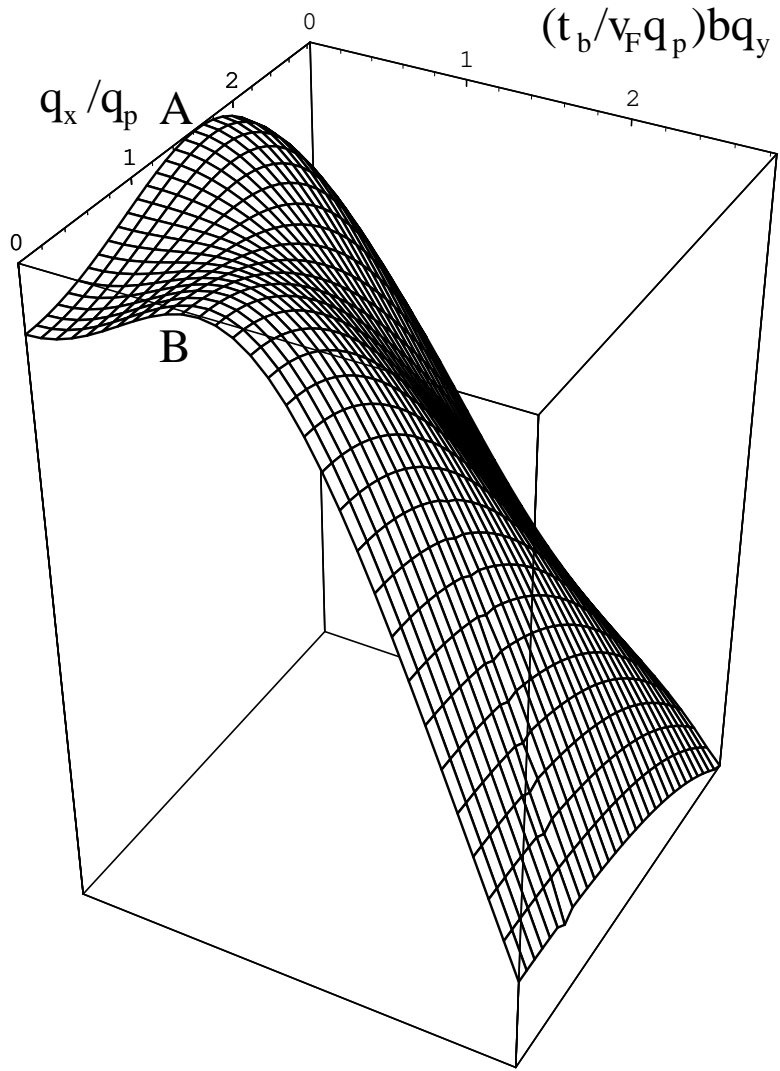
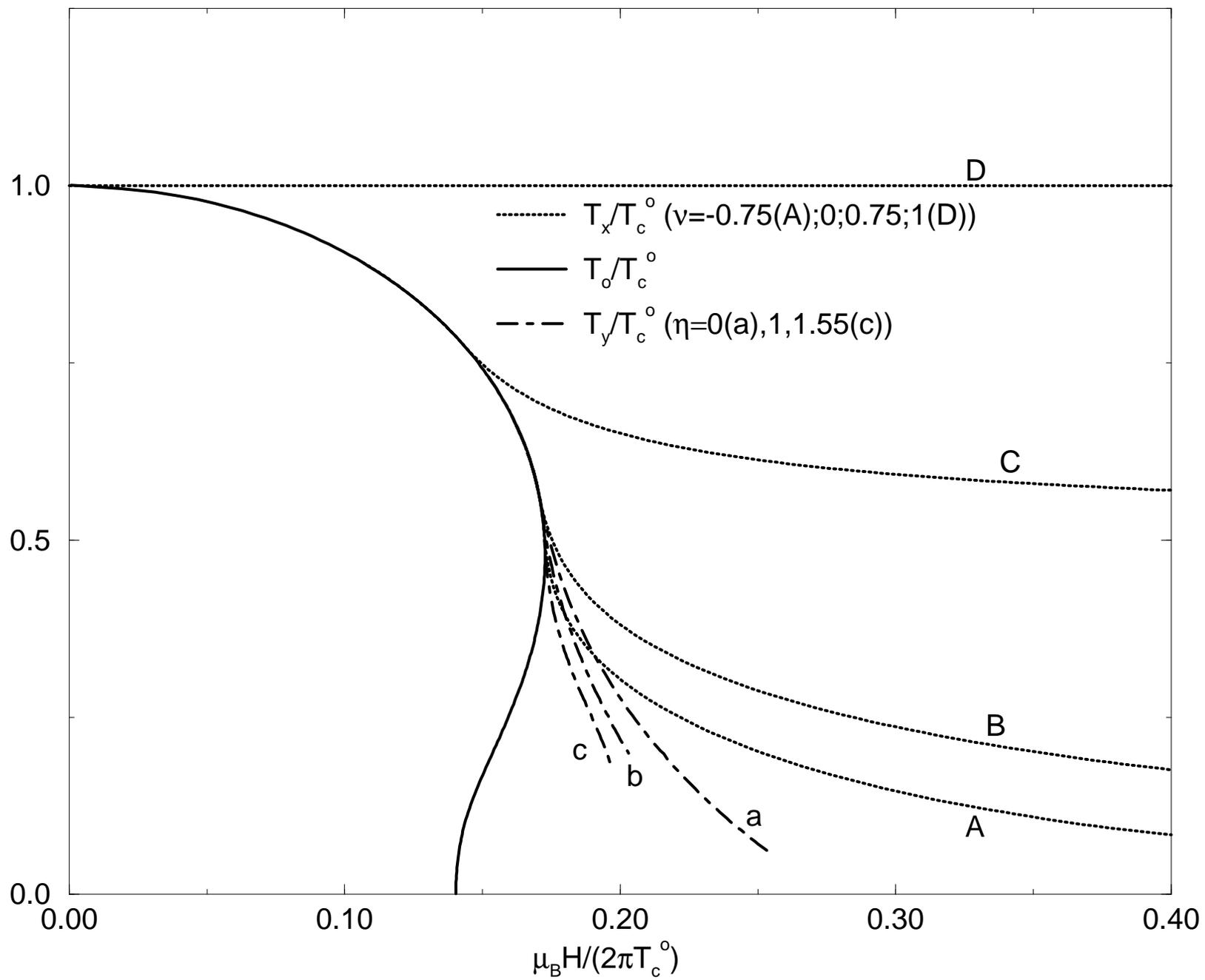
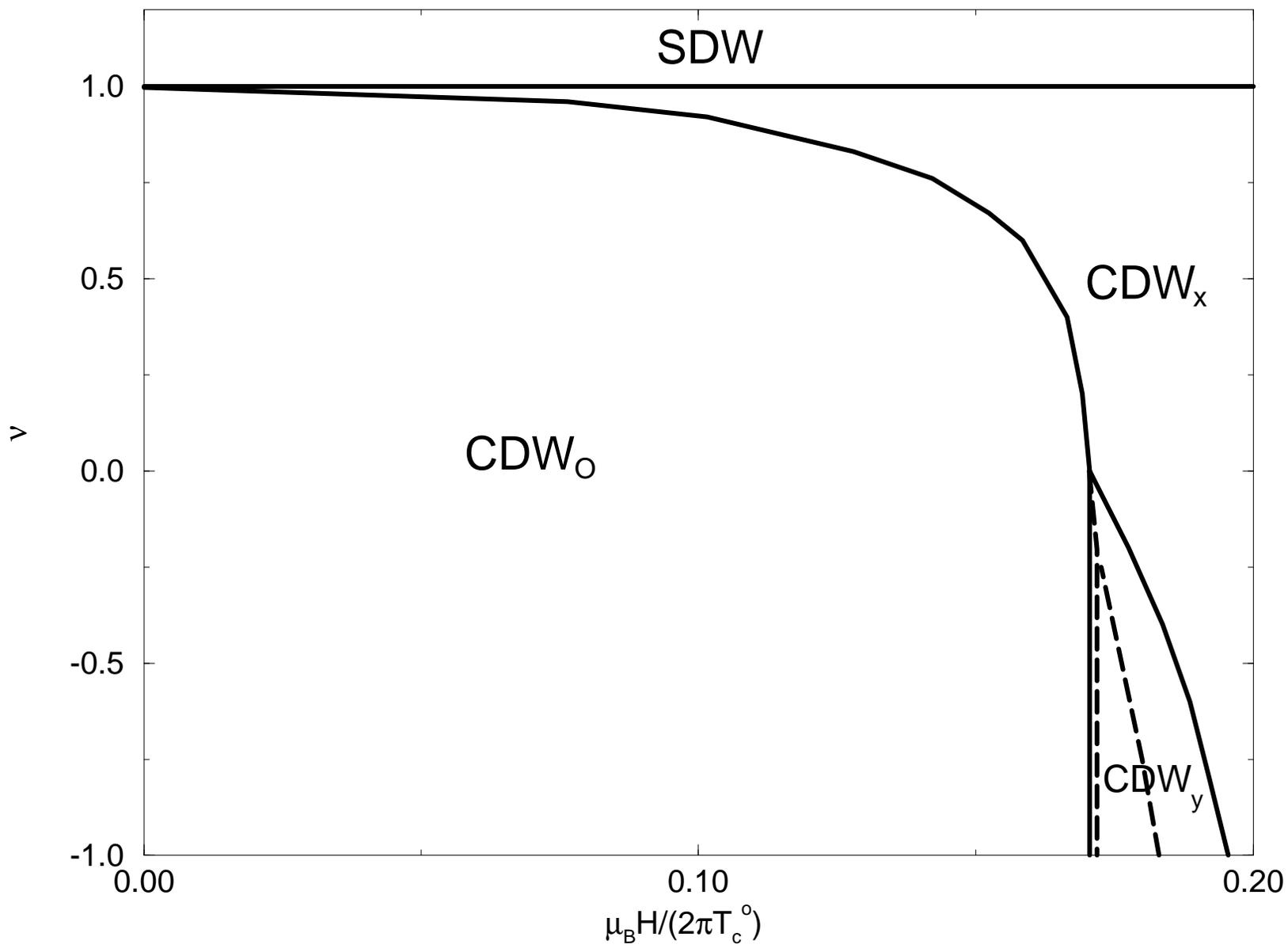
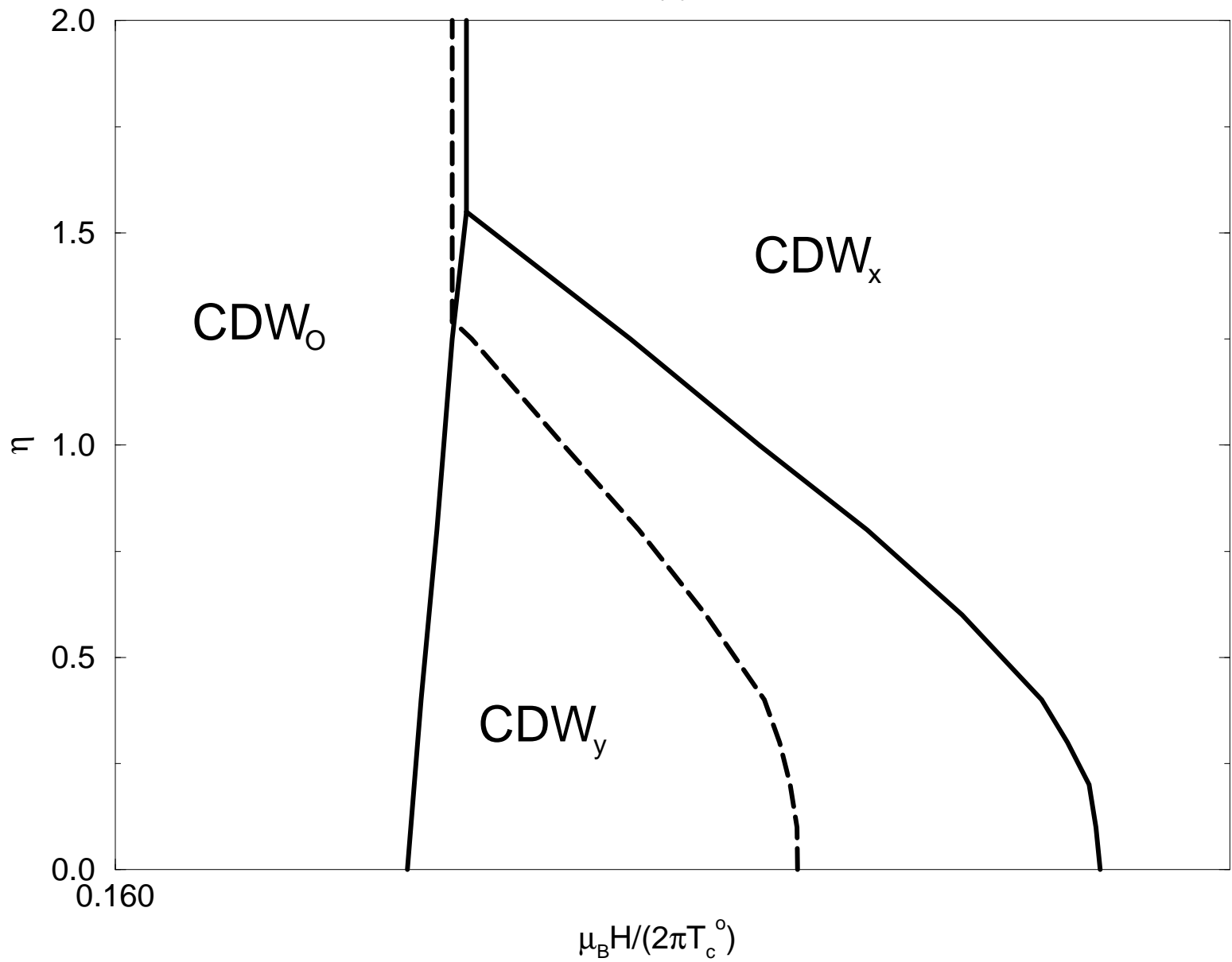
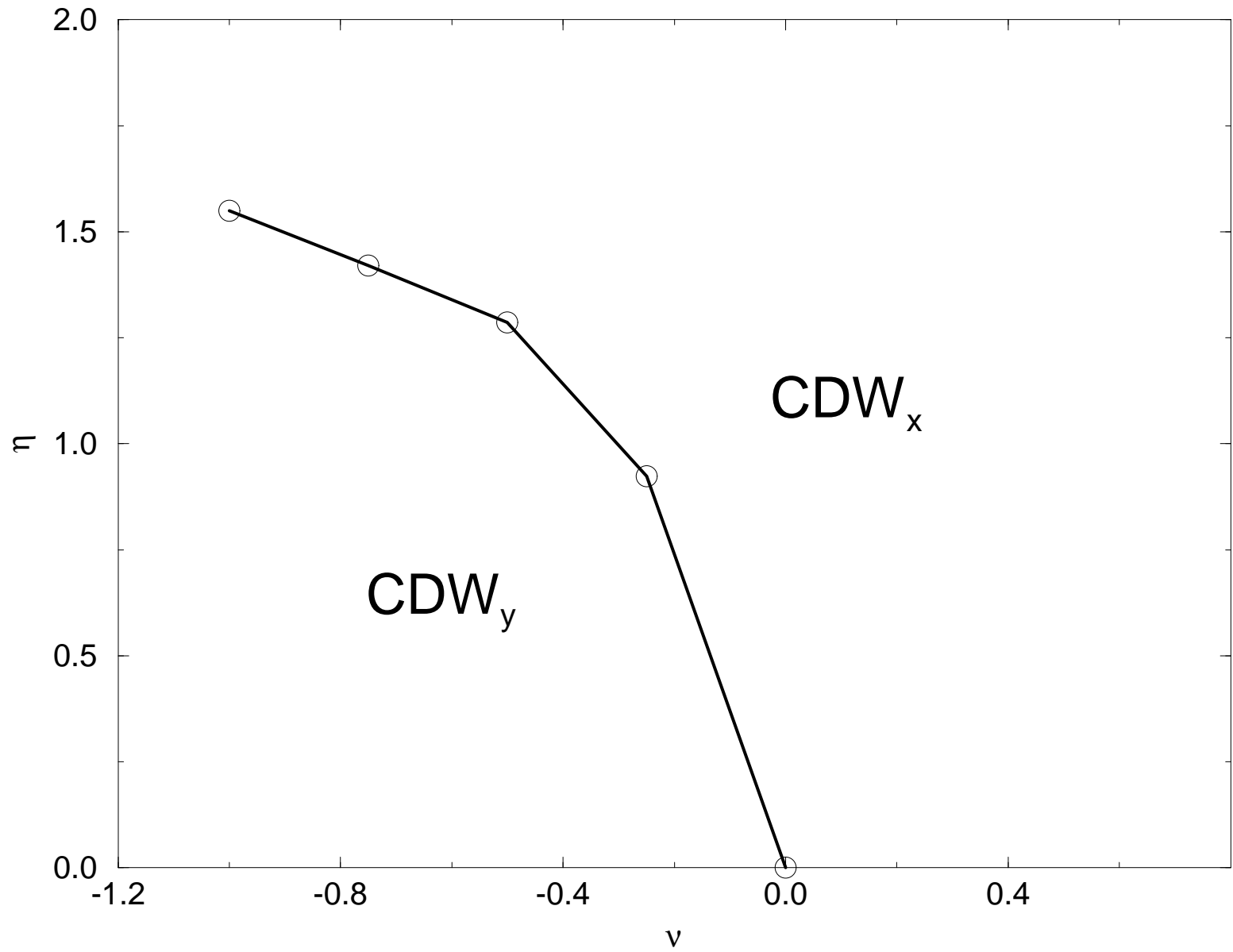
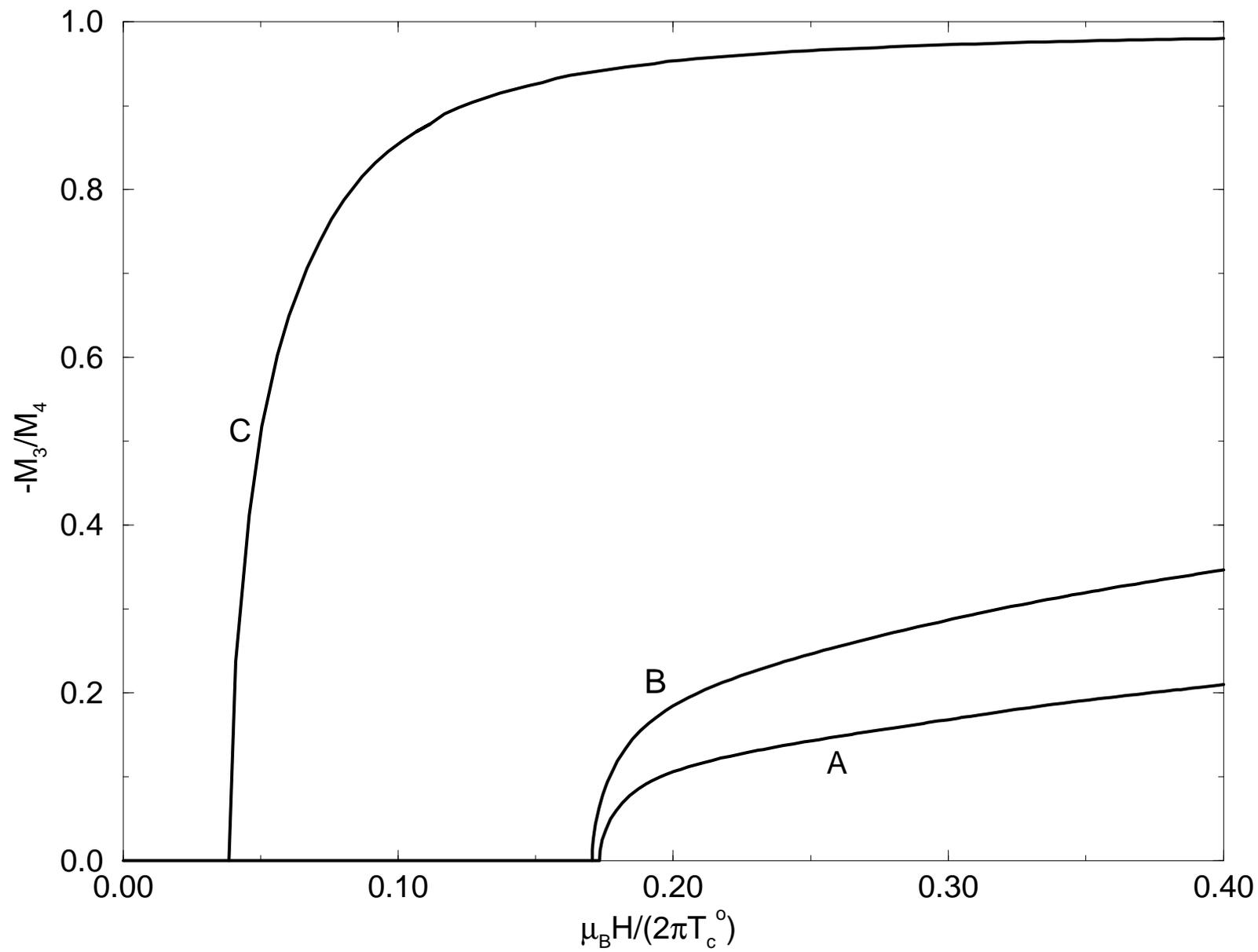
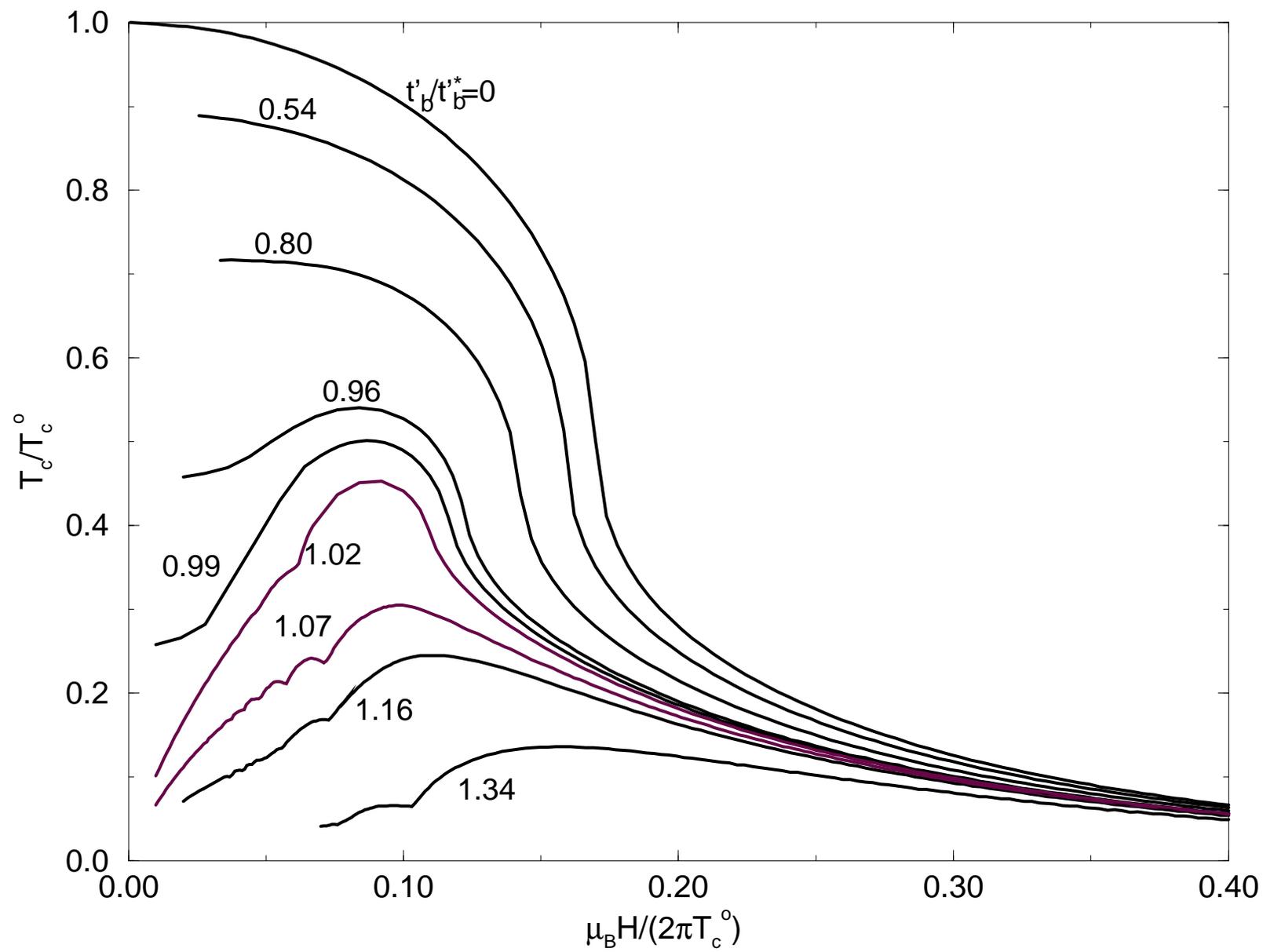
\end{document}